\documentclass[%
 aip,
 amsmath,amssymb,
 reprint,%
]{revtex4-1}

\usepackage{graphicx}
\usepackage{dcolumn}
\usepackage{bm}

\usepackage[utf8]{inputenc}
\usepackage[T1]{fontenc}
\usepackage{mathptmx}
\usepackage{etoolbox}
\usepackage[colorlinks=true, citecolor=blue, linkcolor=blue, urlcolor=blue]{hyperref}

\makeatletter
\def\@email#1#2{%
 \endgroup
 \patchcmd{\titleblock@produce}
  {\frontmatter@RRAPformat}
  {\frontmatter@RRAPformat{\produce@RRAP{*#1\href{mailto:#2}{#2}}}\frontmatter@RRAPformat}
  {}{}
}%
\makeatother
\begin{document}

\preprint{AIP/123-QED}

\title{Revealing Magnon–Photon Coupling in Ultrathin Magnetic Films Using the Derivative-Divide Method}
\author{Kang An}
\author{Zhenhui Hao}
\affiliation{
Key Laboratory of Magnetism and Magnetic Functional Materials
of the Ministry of Education,
Lanzhou University,
Lanzhou 730000,
People's Republic of China}

\author{Yongzhang Shi}
\affiliation{
State Key Laboratory of Crystal Materials,
School of Physics,
Shandong University,
Jinan 250100,
People's Republic of China
}

\author{Yingjie Zhu}
\author{Xiling Li}
\affiliation{
Key Laboratory of Magnetism and Magnetic Functional Materials
of the Ministry of Education,
Lanzhou University,
Lanzhou 730000,
People's Republic of China}

\author{Chi Zhang}
\thanks{Corresponding author: zc@lzu.edu.cn}
\affiliation{
Key Laboratory of Magnetism and Magnetic Functional Materials
of the Ministry of Education,
Lanzhou University,
Lanzhou 730000,
People's Republic of China}

\author{Guozhi Chai}
\thanks{Corresponding author: chaigzh@lzu.edu.cn}
\affiliation{
Key Laboratory of Magnetism and Magnetic Functional Materials
of the Ministry of Education,
Lanzhou University,
Lanzhou 730000,
People's Republic of China}

\date{\today}

\begin{abstract}
Magnon–photon coupling in planar cavities provides a promising platform for integrated cavity magnonics devices.
However, its characterization becomes challenging in ultrathin magnetic films because their weak magnetic response can be obscured by the much stronger microwave background.
Here, we employ the derivative-divide method to extract the field-dependent magnetic response from transmission measurements of split-ring resonators.
The method reveals clear magnon–photon anticrossings that are hardly discernible in conventional transmission spectra and enables quantitative extraction of the coupling strength.
We resolve coupling in yttrium iron garnet films down to 60 nm and extend the approach to conductive CoFeB films down to 5 nm. Our results demonstrate a practical method for characterizing magnon–photon coupling in ultrathin magnetic films, providing an experimental route toward cavity magnonics systems based on ultrathin magnetic films.
\end{abstract}

\maketitle

\section{Introduction}
Cavity magnonics investigates the interaction between collective spin excitations (magnons) and confined microwave photons, giving rise to hybrid cavity magnon polaritons \cite{SolidStatePhys.69.47,SolidStatePhys.70.1,SolidStatePhys.71.117,PhysRep.979.2022}.
Yttrium iron garnet (YIG), owing to its exceptionally low magnetic damping, is widely used for realizing magnon–photon coupling (MPC) \cite{PhysRep.229.1993,JPHYSDAPPLPHYS.43.264002}.
Strong MPC has been extensively demonstrated in three-dimensional microwave cavities, establishing the foundation for cavity magnonics and enabling diverse hybrid phenomena \cite{PhysRevLett.113.156401,NpjQuantumInf.1.15014,NatCommun.6.8914,SciAdv.2.e1501286,NatCommun.8.1368,SciAdv.3.e1603150,PhysRevLett.121.137203,PhysRevLett.125.237201,PhysRevB.106.054425,PhysRevLett.132.137201,PhysRevLett.132.206902,zhang_simultaneous_2025,PhysRevB.112.064416}.
More recently, the development of compact and integrated cavity magnonics has motivated increasing use of planar resonators, which offer reduced mode volumes and greater compatibility with lithographic fabrication \cite{JApplPhys.116.243906,PhysRevLett.123.127202,PhysRevLett.125.147202,PhysRevLett.130.046705,PhysRevLett.132.156901,NatCommun.15.9014,PhysRevLett.132.116701}.

Planar cavities have enabled a broad range of magnon–photon phenomena with YIG spheres, including nonreciprocal coupling, spatial control of the interaction, and long-distance coupling mediated by microwave photons \cite{PhysRevLett.123.127202,PhysRevLett.125.147202,PhysRevAppl.14.014035,ApplPhysLett.119.132403,PhysRevLett.132.156901,PhysRevAppl.11.054023,PhysRevB.101.064404,PhysRevLett.130.046705,PhysRevLett.131.106702}.
Compared with spherical samples, magnetic films provide an extended interaction area with planar resonators and are naturally compatible with lithographically defined structures, making them attractive for integrated cavity magnonics systems.
Early experiments demonstrated MPC between YIG films and superconducting resonators \cite{PhysRevLett.111.127003}, followed by studies involving spin pumping \cite{PhysRevLett.114.227201}, microwave-field-controlled coupling \cite{JPHYSDAPPLPHYS.52.305003}, non-Hermitian photon–magnon responses \cite{NatCommun.15.9014}, and coupling enhancement through resonator engineering \cite{PhysRevAppl.21.034034}.
Subsequent studies have extended planar cavity MPC to various YIG films \cite{JApplPhys.116.243906,SciRep.7.11930,PhysRevB.102.014453,PhysRevAppl.19.014075,NanoLett.23.5055,PhysRevAppl.22.034004,PhysRevLett.132.116701} and metallic magnetic films \cite{PhysRevLett.123.107702,PhysRevLett.123.107701,SciAdv.7.eabe8638}, demonstrating the broad material compatibility of planar cavity magnonics platforms.
Despite these advances, YIG films employed in most planar MPC experiments have historically remained on the micrometer scale, although recent studies have begun to access the sub-100 nm regime using specialized microwave resonators \cite{5.0309552,1mc9-k683}.
Extending quantitative MPC characterization further into nanometer-scale magnetic films, while maintaining sufficient visibility of the weak magnetic response, is therefore important for connecting cavity magnonics with thin-film magnonics and spintronics.

As the thickness of the magnetic film is reduced, the number of participating spins and the total magnetic moment decrease, resulting in a progressively weaker magnon response.
In planar resonators, this weak response is superimposed on a much stronger microwave background dominated by cavity photon modes, making FMR and MPC signatures difficult to distinguish in conventional transmission spectra even when the underlying interaction remains present.
The derivative-divide method, originally developed by Maier-Flaig \textit{et al}. for broadband frequency-domain FMR analysis \cite{RevSciInstrum.89.076101}, suppresses field-independent or slowly field-varying microwave backgrounds while retaining the field-dependent magnetic contribution, thereby improving its visibility without altering the intrinsic sensitivity or frequency resolution of the VNA.
Whether this approach can enable quantitative characterization of MPC in nanometer-thick insulating and metallic magnetic films remains to be established.

Here, we employ the derivative-divide method to characterize MPC in ultrathin magnetic films coupled to a planar split-ring resonator (SRR).
Comparison with conventional transmission measurements shows that the derivative-divide results reveal FMR dispersion and magnon–photon anticrossings that are otherwise difficult to distinguish, enabling quantitative extraction of the coupling strength.
We resolve MPC in YIG films down to 60 nm and further extend the approach to conductive CoFeB films, where coupling remains resolvable down to 5 nm.
These results provide a practical approach for characterizing MPC in ultrathin insulating and metallic magnetic films and a basis for cavity magnonics systems incorporating nanometer-scale magnetic materials.

\section{Experiments and methods} 

\subsection{Experimental Setup and Samples}
\begin{figure}[htbp]
\centering
\includegraphics[width=0.8\linewidth]{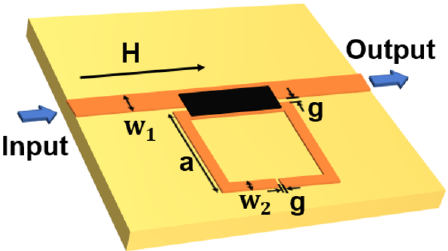}
\caption{Schematic of the SRR and experimental configuration.
The magnetic film is placed with the film side facing downward and in direct contact with the SRR.
The static magnetic field is applied in the film plane and parallel to the microstrip line.}
\label{Fig:1}
\end{figure}

\begin{figure}[htbp]
\centering
\includegraphics[width=0.8 \linewidth]{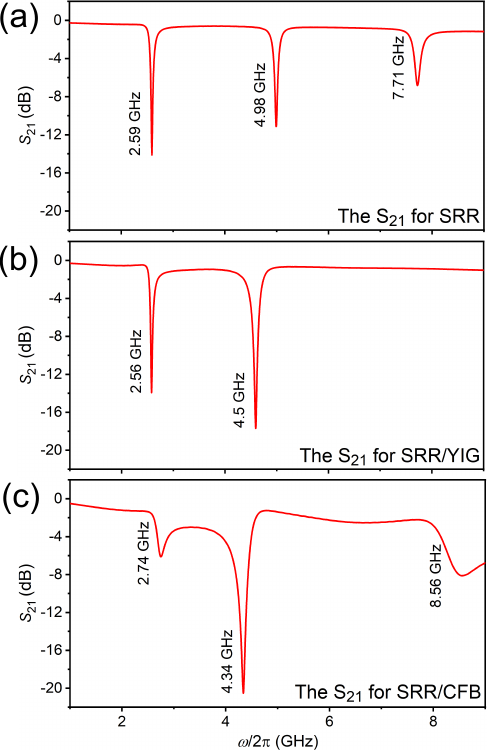}
\caption{Zero-field microwave transmission spectra of (a) the bare SRR, (b) the SRR loaded with a YIG film, and (c) the SRR loaded with a CoFeB film.
The loading of the magnetic films modifies the resonance frequencies and linewidths of the SRR photon modes.}
\label{Fig:2}
\end{figure}

The experimental setup consists of a vector network analyzer (VNA), an SRR, magnetic film samples, and an electromagnet.
As shown in Fig.~\ref{Fig:1}, the SRR is fabricated by optical lithography on a dielectric substrate with a relative permittivity of 3.55 and a thickness of 0.778 mm.
The metallic layers forming the SRR, microstrip line, and ground plane are 35 $\mu$m thick.
The geometrical parameters are $a=10$ mm, $g=0.3$ mm, $w_{1}=1.68$ mm, and $w_{2}=0.65$ mm, as indicated in Fig.~\ref{Fig:1}.
The bare SRR supports three resonance modes at 2.59, 4.98, and 7.71 GHz, as shown in Fig.~\ref{Fig:2}(a).

Both YIG and CoFeB films are investigated.
The YIG films are grown by pulsed laser deposition on gadolinium gallium garnet substrates and have thicknesses of 100, 80, and 60 nm, with identical lateral dimensions of $10~\mathrm{mm}\times5~\mathrm{mm}$.
The CoFeB films are deposited by magnetron sputtering on silicon substrates and have thicknesses of 114.65, 15.92, and 5 nm, with identical lateral dimensions of $5~\mathrm{mm}\times5~\mathrm{mm}$.
Within each material series, the lateral dimensions are fixed while only the film thickness is varied, enabling a direct comparison of the thickness-dependent response.
For all measurements, the samples are placed at the position indicated in Fig.~\ref{Fig:1}.
An in-plane static magnetic field is applied parallel to the microstrip line, and the microwave transmission parameter $S_{21}$ is recorded by the VNA as a function of frequency and applied magnetic field for subsequent analysis.

\subsection{Derivative-Divide Method}
The derivative-divide method was originally developed by Maier-Flaig \textit{et al}. for the analysis of broadband ferromagnetic resonance in the frequency domain \cite{RevSciInstrum.89.076101}.
Here, we employ this method to reveal the weak field-dependent magnetic response that can be obscured by the much stronger microwave background in the SRR transmission spectra.
The method is applied as a numerical post-processing procedure to the complex transmission coefficient $S_{21}(\omega,H)$ measured by the VNA.
For a given magnetic field $H$, the derivative-divide response is calculated from two neighboring spectra as
\begin{equation}
d_D S_{21}(\omega,H)
=
\frac{
S_{21}(\omega,H+\Delta H)
-
S_{21}(\omega,H-\Delta H)
}{
2\Delta H\,S_{21}(\omega,H)
},
\end{equation}
where $\Delta H$ is the magnetic field interval between the central field $H$ and each neighboring spectrum.

The physical origin of the derivative-divide response can be understood from the field dependence of the dynamic magnetic susceptibility $\chi$.
Following the formalism of Ref.~\cite{RevSciInstrum.89.076101}, taking the difference between neighboring spectra largely suppresses the field-independent or slowly field-varying microwave background, while normalization by $S_{21}(\omega,H)$ further reduces the influence of the frequency-dependent transmission background and phase contribution of the measurement chain.
For a sufficiently small field interval, the derivative-divide response can be written as
\begin{equation}
d_D S_{21}
\propto
-i\omega A\frac{\partial\chi}{\partial H}
=
-i\omega A'\frac{\partial\chi}{\partial\omega},
\end{equation}
where $A$ and $A'$ are scaling factors associated with the microwave coupling and the conversion between the field and frequency derivatives, respectively.
Thus, the field-dependent magnetic response is retained, whereas contributions that vary weakly with magnetic field are strongly suppressed.

In the present MPC measurements, the derivative-divide method is used to improve the visibility of the FMR dispersion and hybrid modes that are difficult to distinguish in conventional transmission spectra.
As a numerical post-processing procedure, it does not alter the intrinsic sensitivity, noise floor, or frequency resolution of the VNA, but instead improves the spectral contrast of the magnetic response against the microwave background.
The magnitude of the derivative-divide signal is therefore not interpreted as a direct measure of the coupling strength.
Instead, the MPC strength is quantitatively extracted from the field-dependent dispersion of the resolved hybrid modes using the coupling model discussed below.
Additional details of the derivative-divide data processing are provided in the Supplementary Material.

\section{RESULTS AND DISCUSSION} 
\subsection{Revealing magnon-photon coupling by derivative-divide analysis}
\begin{figure*}[htbp]
\centering
\includegraphics[width=0.8\linewidth]{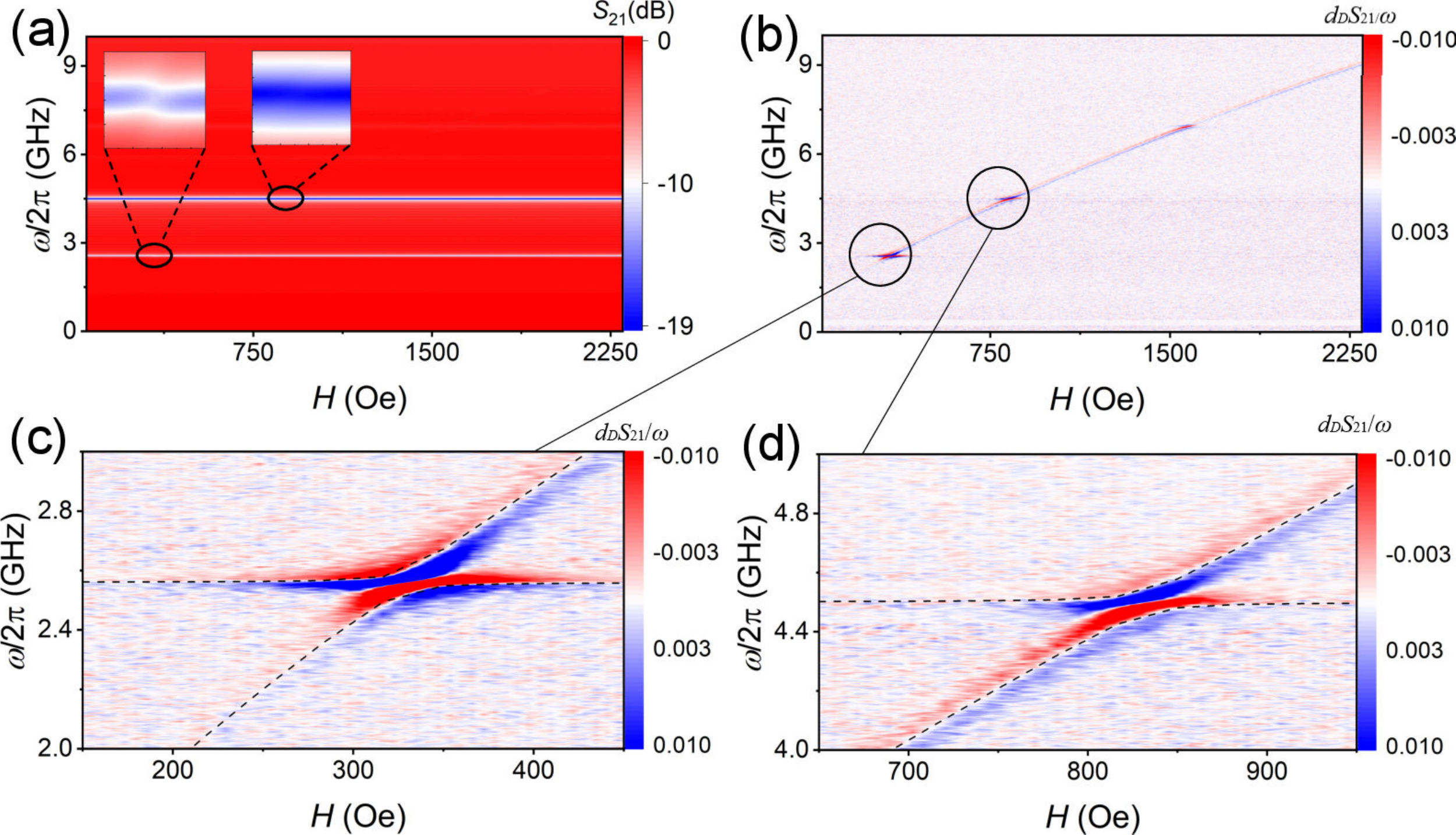}
\caption{Comparison between conventional transmission and derivative-divide analysis for the 100 nm YIG film.
(a) Mapping of the transmission amplitude $S_{21}$ as a function of frequency and applied magnetic field, with the relevant coupling regions shown on an enlarged scale.
(b) Corresponding derivative-divide map, in which the field-dependent FMR response and MPC features become clearly visible.
(c) and (d) Enlarged derivative-divide spectra around the photon modes at 2.56 GHz and 4.5 GHz, respectively.
Fitting the dispersions of coupling modes with Eq.~\ref{3} gives coupling strengths $\kappa_q/2\pi$ of 50 MHz and 54 MHz, respectively.}
\label{Fig:3}
\end{figure*}

We first use the 100 nm YIG film to examine how the derivative-divide analysis changes the visibility of MPC in the SRR transmission spectra.
When the YIG film is loaded onto the SRR, the two photon modes considered here are located at 2.56 GHz and 4.5 GHz, as shown in Fig.~\ref{Fig:2}(b).
Figure~\ref{Fig:3}(a) shows the conventional transmission parameter $S_{21}$ as a function of frequency $\omega/2\pi$ and applied magnetic field $H$.
The spectrum is dominated by the SRR photon modes, whereas the field-dependent FMR response of the YIG film is much weaker and is difficult to distinguish directly.
As highlighted in the enlarged regions of Fig.~\ref{Fig:3}(a), only a weak perturbation can be discerned for the photon mode with a relatively narrow linewidth, while the interaction is barely visible for the broader photon mode.
This comparison indicates that the visibility of MPC in the conventional transmission spectrum depends not only on the underlying coupling, but also on the relative strength of the magnetic response, the microwave background, and the linewidth of the photon mode.

The same experimental data processed using the derivative-divide method are shown in Fig.~\ref{Fig:3}(b).
In contrast to the conventional transmission map, the field-dependent FMR dispersion becomes clearly visible after the weakly field-dependent SRR background is suppressed.
More importantly, distinct deviations from the uncoupled FMR dispersion appear when the magnon mode approaches the photon modes, revealing the upper and lower hybrid branches associated with MPC.
The derivative-divide procedure, therefore, does not create an additional mode splitting but makes the field-dependent hybridization that is difficult to resolve in the original transmission spectrum spectrally accessible.
For visualization, the real part of $d_D S_{21}/\omega$, which is proportional to the frequency derivative of the magnetic susceptibility, is plotted using a red-blue color scale.
Under the phase convention used here, the red and blue regions represent opposite signs of the derivative response, while the boundary between them corresponds to $d_D S_{21}=0$.
According to the relation between $d_D S_{21}$ and $d\chi/d\omega$ discussed in Sec.~II.B, this zero-crossing traces the resonance frequency and provides a convenient reference for identifying the FMR dispersion and the hybrid branches.

The two coupling regions are shown in greater detail in Figs.~\ref{Fig:3}(c) and \ref{Fig:3}(d).
Near resonance, the uncoupled magnon and photon modes hybridize and form two branches with frequencies $\omega_{+}$ and $\omega_{-}$.
The coupling strength $\kappa_q$ is quantitatively extracted by fitting the experimentally determined frequencies of coupling modes using \cite{PhysRevLett.113.156401}
\begin{equation}\label{3}
\omega_\pm
=
\frac{1}{2}(\omega_{c}+\omega_{m})
\pm
\frac{1}{2}
\sqrt{(\omega_{c}-\omega_{m})^2+4\kappa_q^2},
\end{equation}
where $\omega_c$ and $\omega_m$ are the uncoupled photon and magnon frequencies, respectively.
At zero detuning, where $\omega_c=\omega_m$, the coupling strength satisfies $\kappa_q/2\pi=|\omega_+-\omega_-|/4\pi$.
Fitting the coupling regions at 2.56 GHz and 4.5 GHz yields $\kappa_q/2\pi=50$ MHz and 54 MHz, respectively.
The observation of both hybrid branches and their agreement with the dispersion of coupling modes demonstrate that the derivative-divide analysis enables quantitative extraction of MPC even when the corresponding magnetic features have poor visibility in the conventional $S_{21}$ spectrum.

The comparison in Fig.~\ref{Fig:3} also clarifies the role of the derivative-divide method in the present measurements.
Its principal advantage is the suppression of the strong microwave background and the resulting enhancement of the spectral visibility of the field-dependent magnetic response, rather than an improvement of the intrinsic sensitivity or resolution of the VNA.
The 100 nm YIG film therefore provides a reference system for evaluating MPC from the derivative-divide spectra.
We next use the same analysis to determine how the observable MPC evolves as the YIG film thickness is reduced.

\subsection{Thickness-dependent MPC in YIG films}

\begin{figure}[htbp]
\centering
\includegraphics[width=0.8\linewidth]{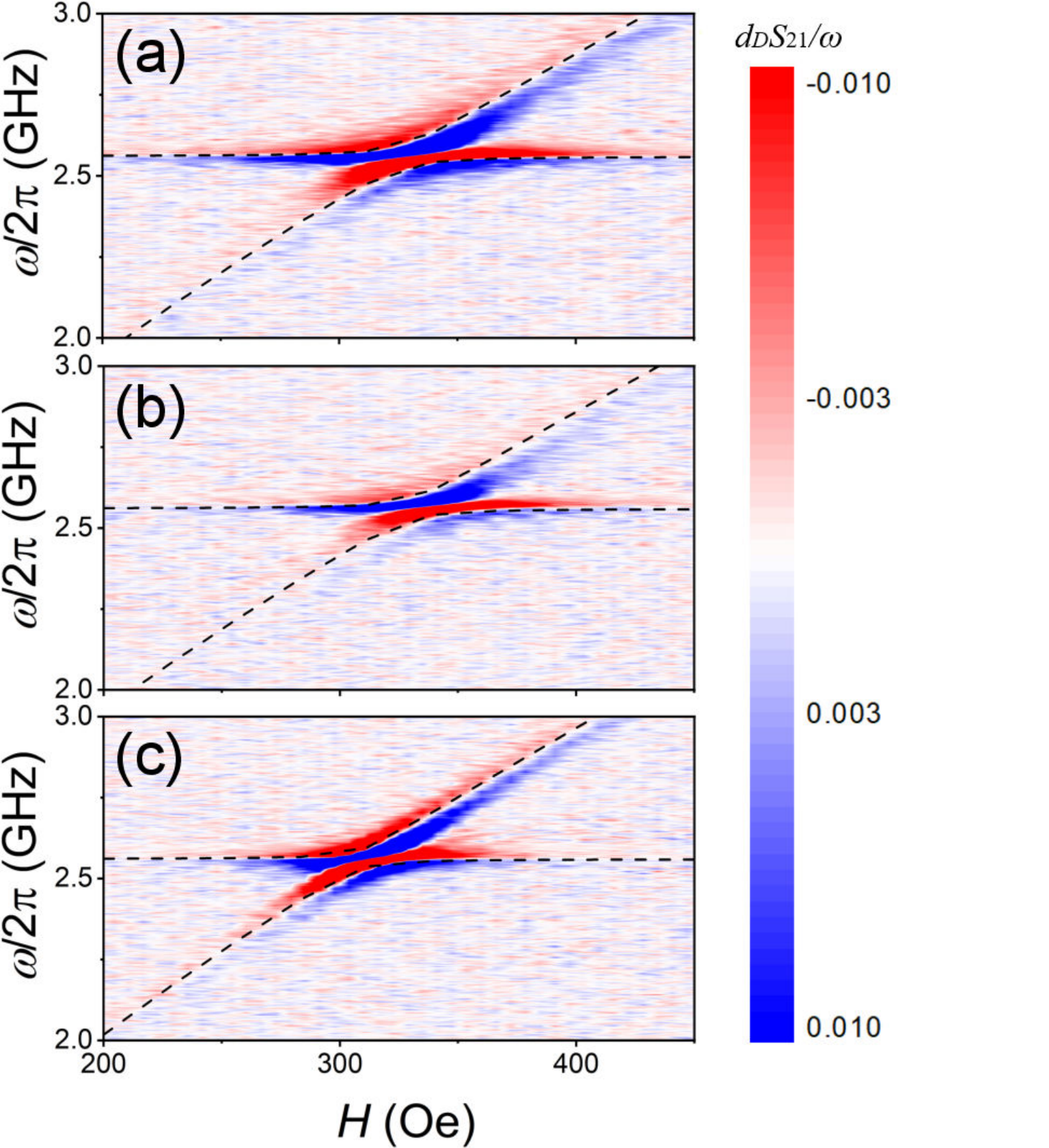}
\caption{Thickness-dependent MPC in YIG films.
Derivative-divide spectra around the first photon mode for YIG films with thicknesses of (a) 100 nm, (b) 80 nm, and (c) 60 nm.
The corresponding coupling strengths $\kappa_q/2\pi$, extracted using Eq.~\ref{3}, are 50 MHz, 47 MHz, and 40 MHz, respectively.}
\label{Fig:4}
\end{figure}

Having established the ability of the derivative-divide method to reveal MPC in the 100 nm YIG film, we next investigate how the coupling evolves as the YIG thickness is reduced.
The 100 nm, 80 nm, and 60 nm YIG films have identical lateral dimensions and are measured under the same sample configuration, allowing the effect of film thickness to be examined directly.
Figure~\ref{Fig:4} compares the derivative-divide spectra around the first photon mode for the three YIG films.
Clear field-dependent hybridization is observed for all three thicknesses, including the 60 nm film, demonstrating that the magnon-photon anticrossing remains resolvable as the magnetic volume is substantially reduced.

The dispersions of coupling modes are fitted using Eq.~\ref{3}, yielding coupling strengths $\kappa_q/2\pi$ of 50 MHz, 47 MHz, and 40 MHz for the 100 nm, 80 nm, and 60 nm films, respectively.
Thus, $\kappa_q$ decreases monotonically as the YIG thickness is reduced.
For collective MPC, the coupling strength is expected to scale approximately with the square root of the number of participating spins, $\kappa_q\propto\sqrt{N}$ \cite{PhysRep.979.2022}.
Because the three YIG samples have the same lateral dimensions and are coupled to the same SRR photon mode, the number of participating spins is approximately proportional to the film thickness, giving $\kappa_q\propto\sqrt{t}$ to first order.
Using the measured value of 50 MHz for the 100 nm film as a reference, this scaling gives approximately 44.7 MHz and 38.7 MHz for the 80 nm and 60 nm films, respectively, which are close to the experimentally extracted values of 47 MHz and 40 MHz.
The observed thickness dependence is therefore consistent with the reduction in the number of participating spins as the magnetic film becomes thinner.

To place the present thickness range in the context of previous experiments of YIG films, Table~\ref{tab:YIGcomparison} summarizes representative studies using planar microwave resonators.
Earlier experiments predominantly employed YIG films with thicknesses ranging from several to several tens of micrometers \cite{PhysRevB.102.014453,SciRep.7.11930,PhysRevAppl.19.014075,JPHYSDAPPLPHYS.52.305003,PhysRevAppl.21.034034}.
More recent studies have extended experiments into the sub-100-nm regime.
Grishin and Nikitov demonstrated cavity-enhanced magnetic dynamics using a 100 nm YIG film integrated with a half-wave microstrip resonator \cite{5.0309552}, while Zanichelli \textit{et al}. achieved strong MPC in a 75 nm YIG film using a loop-gap resonator together with field-differential spectroscopy \cite{1mc9-k683}.
The present derivative-divide measurements further resolve MPC in a 60 nm YIG film using an SRR.

\begin{table*}[htbp]
\caption{Representative experimental studies of MPC and cavity-enhanced magnetic dynamics in YIG thin films using planar microwave resonators.}
\label{tab:YIGcomparison}
\centering
\begin{tabular}{lcccc}
\hline\hline
Magnetic film & Thickness & Resonator & Spectral analysis & Reference \\
\hline
La:YIG & 83 $\mu$m & Stripline resonator & Transmission & \cite{PhysRevB.102.014453} \\
YIG & 25 $\mu$m & Inverted SRR & Transmission & \cite{SciRep.7.11930} \\
YIG & 5 $\mu$m & Metamaterial planar resonator & Transmission & \cite{PhysRevAppl.19.014075} \\
YIG & 4 $\mu$m & SRR & Transmission & \cite{JPHYSDAPPLPHYS.52.305003} \\
YIG & 2 $\mu$m & Double-SRR & Transmission & \cite{PhysRevAppl.21.034034} \\
YIG & 100 nm & Half-wave microstrip resonator & Transmission / nonlinear response & \cite{5.0309552} \\
YIG & 75 nm & Loop-gap resonator & Field-differential spectroscopy & \cite{1mc9-k683} \\
YIG & 60 nm & SRR & Derivative-divide & This work \\
\hline\hline
\end{tabular}
\end{table*}

Because the resonator geometry, resonance frequency, microwave-field distribution, filling factor, sample dimensions, and spectral analysis differ among these studies, the film thicknesses listed in Table~\ref{tab:YIGcomparison} should not be interpreted as a direct ranking of instrumental sensitivity.
Rather, the comparison shows that MPC experiments with YIG have recently progressed from predominantly micrometer into the sub-100-nm regime.
Within this emerging regime, the present results demonstrate that derivative-divide analysis enables the field-dependent dispersions of coupling modes to remain quantitatively resolvable in a 60 nm YIG film using an SRR.

Most importantly, even for the 60 nm YIG film, the upper and lower hybrid branches remain sufficiently resolved in the derivative-divide spectrum to permit quantitative determination of $\kappa_q$.
This result extends the characterization of SRR-mediated MPC into the tens-of-nanometers thickness range without attributing the improvement to a change in the intrinsic sensitivity of the VNA.
Instead, the derivative-divide analysis enables the weak field-dependent hybrid response to be distinguished from the much stronger microwave background.
Having demonstrated this capability for low-damping insulating YIG films, we next examine whether the same approach can be extended to conductive CoFeB films with substantially smaller thicknesses.

\subsection{Extension to ultrathin CoFeB films}

We next extend the derivative-divide analysis from insulating YIG to conductive CoFeB films.
As shown in Fig.~\ref{Fig:2}(c), loading the CoFeB film modifies the SRR photon modes from the bare-cavity resonances at 2.59, 4.98, and 7.71 GHz to approximately 2.74, 4.34, and 8.56 GHz, respectively.
The pronounced frequency shifts indicate that the conductive CoFeB film introduces appreciable electromagnetic loading to the SRR.
Control measurements with the bare GGG and Si substrates show only minor changes in the SRR resonances, indicating that the more pronounced modifications observed after loading the magnetic films originate primarily from the films rather than from the substrates, as shown in the Supplementary Material.
In the following measurements, we focus on the photon mode near 8.56 GHz, for which the field-dependent magnetic response and its interaction with the CoFeB FMR can be clearly followed after derivative-divide processing.

\begin{figure}[htbp]
\centering
\includegraphics[width=0.9\linewidth]{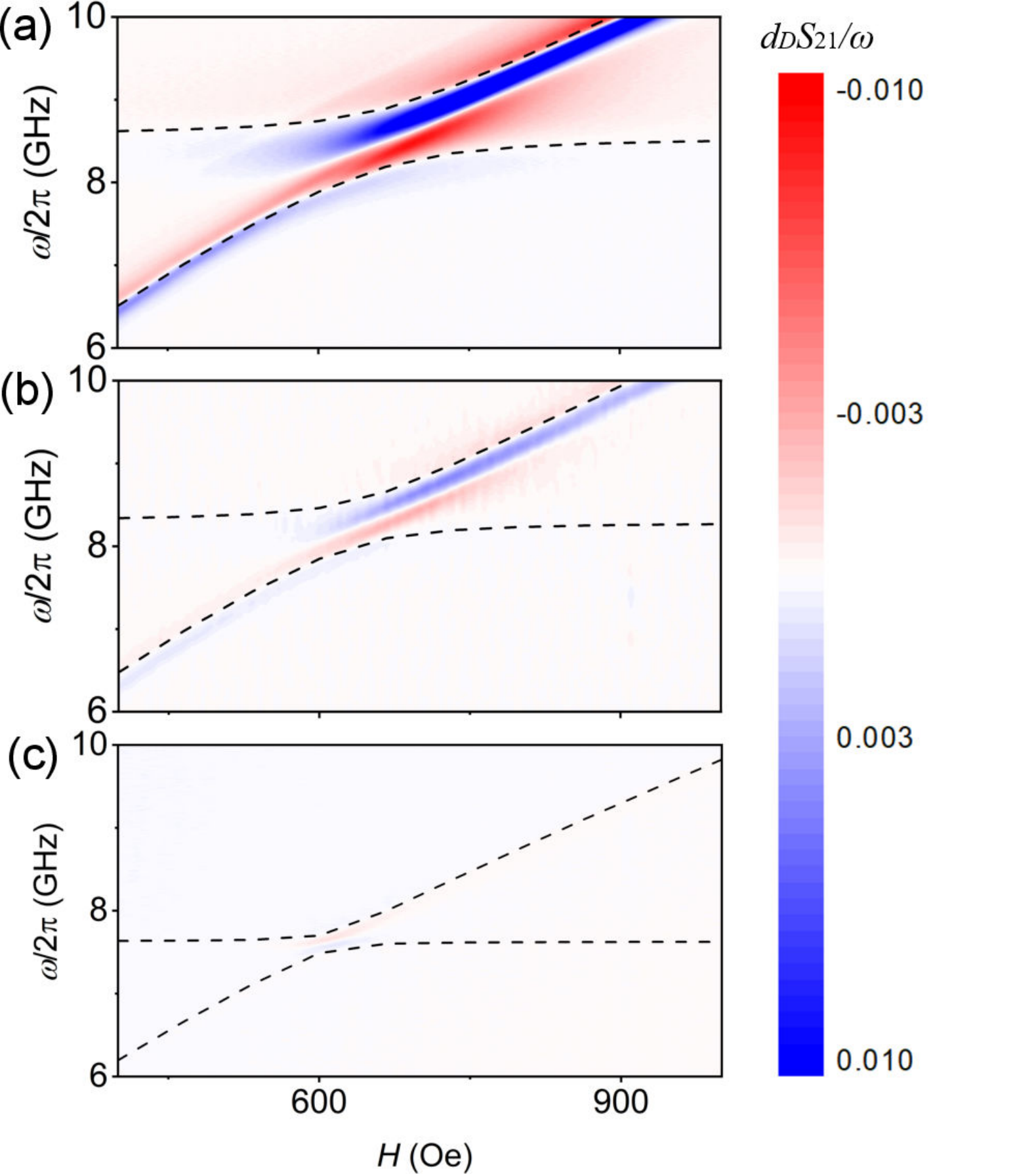}
\caption{Thickness-dependent MPC in CoFeB films.
Derivative-divide spectra around the SRR photon mode near 8.56 GHz for CoFeB films with thicknesses of (a) 114.65 nm, (b) 15.92 nm, and (c) 5 nm.
The coupling strengths $\kappa_q/2\pi$ extracted from the dispersions of coupling modes using Eq.~\ref{3} are 350 MHz, 270 MHz, and 100 MHz, respectively.}
\label{Fig:5}
\end{figure}

Figure~\ref{Fig:5} shows the derivative-divide spectra for CoFeB films with thicknesses of 114.65 nm, 15.92 nm, and 5 nm.
For all three samples, the field-dependent CoFeB resonance approaches the SRR photon mode and produces characteristic dispersions of coupling modes.
The coupling strengths are extracted by fitting the resolved branches using Eq.~\ref{3}.
The resulting values of $\kappa_q/2\pi$ are 350 MHz, 270 MHz, and 100 MHz for the 114.65 nm, 15.92 nm, and 5 nm CoFeB films, respectively.

The extracted $\kappa_q$ decreases as the CoFeB film becomes thinner, qualitatively consistent with the reduction in magnetic volume and the number of participating spins.
Unlike the YIG series, however, the thickness dependence of the CoFeB coupling should not be interpreted solely in terms of the collective spin scaling $\kappa_q\propto\sqrt{N}$.
Because CoFeB is electrically conductive, changing its thickness can also modify the electromagnetic loading of the SRR, the local microwave-field distribution, and the spatial overlap between the photon and magnon modes.
These effects can contribute simultaneously to the measured hybridization and make a simple thickness scaling inappropriate for the present CoFeB series.

Nevertheless, the derivative-divide spectra show that the dispersions of coupling modes remain experimentally resolvable as the CoFeB thickness is reduced.
Even for the 5 nm film, a field-dependent deviation from the uncoupled magnon dispersion remains visible around the photon mode, allowing a coupling strength of $\kappa_q/2\pi=100$ MHz to be extracted.
Under the present experimental conditions, 5 nm is the thinnest CoFeB film for which the MPC feature can be reliably resolved.
For still thinner samples, the field-dependent magnetic response becomes too weak relative to the remaining microwave background for reliable determination of the dispersions of coupling modes.

It is also important to distinguish the magnitude of the derivative-divide response from the coupling strength $\kappa_q$.
The spectral contrast of $d_D S_{21}$ depends not only on the MPC, but also on the magnetic susceptibility, linewidth, microwave loading, and mode overlap.
Therefore, a stronger derivative-divide signal does not by itself imply a larger $\kappa_q$.
In the present analysis, $\kappa_q$ is determined exclusively from the field-dependent dispersions using Eq.~\ref{3}.

For the same reason, we do not make a direct quantitative comparison between the values of $\kappa_q$ obtained for YIG and CoFeB.
The two material systems differ in sample area, magnetic damping, saturation magnetization, photon mode, electromagnetic loading, and spatial overlap with the SRR microwave field.
The significance of the CoFeB measurements is therefore not that the extracted coupling strengths are larger than those obtained for YIG, but that MPC remains quantitatively resolvable in a conductive magnetic film with a thickness as small as 5 nm.
Together with the 60 nm YIG results, this demonstrates that derivative-divide analysis provides a practical means of characterizing MPC in both insulating and conductive magnetic films approaching the nanometer thickness regime.

\section{CONCLUSION}
In summary, we have investigated MPC in ultrathin magnetic films using an SRR together with derivative-divide analysis.
By suppressing the strong microwave background that obscures the weak field-dependent magnetic response in conventional transmission spectra, the derivative-divide method reveals the FMR dispersion and branches of coupling modes and enables the coupling strength $\kappa_q$ to be quantitatively extracted from their field-dependent dispersion.
For YIG films with thicknesses of 100, 80, and 60 nm, the extracted $\kappa_q/2\pi$ decreases from 50 to 47 and 40 MHz, respectively, consistent with the reduction in the number of participating spins as the magnetic volume decreases.
The analysis is further extended to conductive CoFeB films, for which MPC remains resolvable down to a thickness of 5 nm, with $\kappa_q/2\pi=100$ MHz extracted from the dispersions of coupling modes.
Importantly, the derivative-divide procedure improves the spectral visibility of the magnetic response rather than the intrinsic sensitivity or frequency resolution of the VNA, and the magnitude of the derivative-divide signal itself is not used as a measure of the coupling strength.
These results establish derivative-divide analysis as a practical approach for quantitatively characterizing MPC in nanometer-scale insulating and conductive magnetic films and provide an experimental route for extending cavity magnonics toward ultrathin magnetic materials.

\section*{SUPPLEMENTARY MATERIAL}
See the Supplementary Material for substrate control measurements using bare GGG and Si substrates and additional details of the derivative-divide data-processing procedure.

\section*{acknowledgments}
This work is supported by the National Natural Science Foundation of China (NSFC) (Nos. 52471200, 12174165 and 52201219) and Scientific Research Innovation Capability Support Project for Young Faculty (ZYGXQNJSKYCXNLZCXM-I19).

\section*{Data Availability Statement}
The data that support the findings of this study are available from the corresponding authors upon reasonable request.

\bibliography{aipsamp}

\end{document}